\documentclass[prd,superscriptaddress,amsfonts,amssymb,amsmath,showpacs,twocolumn,10 pt,superscriptaddress,nofootinbib]{revtex4-2}
\usepackage{bm}
\usepackage{amsfonts}
\usepackage{latexsym}
\usepackage[latin1]{inputenc}
\usepackage{graphicx}
\usepackage{amsmath,eqnarray}
\usepackage{palatino}
\usepackage{mathpazo}
\usepackage{textcomp}
\usepackage{float}
\usepackage{subcaption}
\usepackage{booktabs}
\usepackage{dcolumn}
\usepackage{hyperref}
\usepackage{mathtools}
\hypersetup{colorlinks,citecolor=red}
\usepackage{footnote}
\usepackage{amsmath}
\usepackage{xcolor}
\usepackage{orcidlink}
\usepackage{ragged2e}
\usepackage{amsmath}
\usepackage{listings}
\usepackage{graphicx}
\usepackage{float}
\def\jnl@style{\it}
\def\aaref@jnl#1{{\jnl@style#1}}

\def\aaref@jnl#1{{\jnl@style#1}}

\def\aj{\aaref@jnl{AJ}}                   % Astronomical Journal
\def\apj{\aaref@jnl{ApJ}}                 % Astrophysical Journal
\def\apjl{\aaref@jnl{ApJ}}                % Astrophysical Journal, Letters
\def\apjs{\aaref@jnl{ApJS}}               % Astrophysical Journal, Supplement
\def\apss{\aaref@jnl{Ap\&SS}}             % Astrophysics and Space Science
\def\aap{\aaref@jnl{A\&A}}                % Astronomy and Astrophysics
\def\aapr{\aaref@jnl{A\&A~Rev.}}          % Astronomy and Astrophysics Reviews
\def\aaps{\aaref@jnl{A\&AS}}              % Astronomy and Astrophysics, Supplement
\def\mnras{\aaref@jnl{Mon.~Not.~Roy.~Astron.~Soc.}}             % Monthly Notices of the RAS
\def\prd{\aaref@jnl{Phys.~Rev.~D}}        % Physical Review D
\def\prc{\aaref@jnl{Phys.~Rev.~C}}  % Physical Review C
\def\prl{\aaref@jnl{Phys.~Rev.~Lett.}}    % Physical Review Letters
\def\qjras{\aaref@jnl{QJRAS}}             % Quarterly Journal of the RAS
\def\skytel{\aaref@jnl{S\&T}}             % Sky and Telescope
\def\ssr{\aaref@jnl{Space~Sci.~Rev.}}     % Space Science Reviews
\def\zap{\aaref@jnl{ZAp}}                 % Zeitschrift fuer Astrophysik
\def\nat{\aaref@jnl{Nature}}              % Nature
\def\aplett{\aaref@jnl{Astrophys.~Lett.}} % Astrophysics Letters
\def\apspr{\aaref@jnl{Astrophys.~Space~Phys.~Res.}} % Astrophysics Space Physics Research
\def\physrep{\aaref@jnl{Phys.~Rep.}}      % Physics Reports
\def\physscr{\aaref@jnl{Phys.~Scr}}       % Physica Scripta
\def\commat{\aaref@jnl{Comm.~Math.~Phys.}}              % Communications in Mathematical Physics
\def\science{\aaref@jnl{Science}}               % Science
\def\cqg{\aaref@jnl{Classical Quant.~Grav.}}            % Classical and Quantum Gravity
\def\jpcs{\aaref@jnl{JPCS}}                                     % Journal of Physics Conference Series
\def\ijmpd{\aaref@jnl{Int.~J.~Mod.~Phys.~D}}                    % International Journal of Modern Physics D
\def\grg{\aaref@jnl{Gen.~Relat.~Gravit.}}               % General Relativity and Gravitation
\def\rpp{\aaref@jnl{Rep.~Prog.~Phys.}}          % Reports on Progress in Physics
\def\npa{\aaref@jnl{Nucl.~Phys.~A}}        % Nuclear Physics A
\def\lrr{\aaref@jnl{Living Rev.~Rel.}}                   % Living reviews in relativity
\def\jcap{\aaref@jnl{J.~Cosmology Astropart.~Phys.}}    % Journal of cosmology and astroparticle physics
\def\rmp{\aaref@jnl{Rev.~Mod.~Phys.}}   %Reviews of modern physics
\def\epjc{\aaref@jnl{Eur.~Phys.~J.~C}}

\allowdisplaybreaks[1]
\begin{document}

\color{black}

\author{Gaurav N. Gadbail\orcidlink{0000-0003-0684-9702}}
\email{gauravgadbail6@gmail.com}
\affiliation{Faculty of Symbiotic Systems Science, Fukushima University, Fukushima 960-1296, Japan.}

\author{Kazuharu Bamba \orcidlink{0000-0001-9720-8817}}
\email{bamba@sss.fukushima-u.ac.jp}
\affiliation{Faculty of Symbiotic Systems Science, Fukushima University, Fukushima 960-1296, Japan.}

\title{A model-independent measurement of the Hubble constant from gravitational-wave standard sirens and electromagnetic observations}

%\date{\today}
\begin{abstract}
The Hubble tension is one of the most significant challenges in modern cosmology. Developing new approaches to estimate the Hubble constant is therefore crucial, and in this work, we employ a Gaussian process, a fully model-independent method that relies solely on observational data. To determine the Hubble constant, we use not only electromagnetic observations but also include gravitational-wave standard siren data from GWTC3. Our measurements of the Hubble constant are strongly consistent with the SH0ES result, with tensions less than $2\sigma$, indicating no statistically significant discrepancy. This approach quantifies the impact of gravitational-wave data on the determination of the Hubble constant, examines its consistency with electromagnetic measurements, and explores its potential role in addressing the Hubble tension.

%This inclusion improves the results as compared to previous studies and may provide new clues to fix the Hubble tension. 

% \textbf{Keywords:} Hubble constant; Observational cosmology; Gravitational wave

\end{abstract}

\maketitle

% \date{\today}

\section{Introduction}
The Hubble parameter is one of the most important cosmological parameters, as it describes the expansion history of the universe. The value of the Hubble parameter at $z=0$, known as the Hubble constant ($H_0$), characterizes the present expansion rate of the universe. Recently, the two most powerful observational probes are Type Ia supernovae (SNe Ia) and the cosmic microwave background (CMB), which measure the $H_0$ value. The SNe Ia observations determine the Hubble constant $H_0 = 73.04 \pm 1.04$ km s$^{-1}$ Mpc$^{-1}$ through the local distance ladder, providing a direct measurement of the present expansion rate of the universe \cite{Riess/2022}. This direct measurement is achieved by calibrating the intrinsic brightness of SNe Ia using nearby Cepheid variable stars, known for their consistent luminosity, thus providing a standard candle for distance estimation. On the other hand, the CMB provides a snapshot of the early universe, from which $H_0$ is inferred indirectly via an inverse distance ladder. This relies on the fluctuations observed in the CMB and their dependence on the assumed cosmological model, most commonly the $\Lambda$CDM model. The value inferred from Planck CMB measurements is $H_0 = 67.4 \pm 0.5$ km s$^{-1}$ Mpc$^{-1}$ \cite{Aghanim/2020}.
%which is commonly referred to as the early-Universe determination. In contrast, the SH0ES collaboration estimates $H_0 = 73.04 \pm 1.04$ km s$^{-1}$ Mpc$^{-1}$, representing a late-time observational determination. 
These early- and late-time measurements of $H_0$ are in significant disagreement with each other, and the tension between these measurements is approximately $5\sigma-6\sigma$, making it difficult to explain as a mere statistical fluke. This discrepancy is known as the Hubble tension. Should this tension persist, it could potentially challenge the standard cosmological model, which serves as the foundation for our understanding of the structure and history of the universe. One possible implication of this tension is the need for new physical mechanisms or extensions to the standard model, potentially involving previously unknown aspects of dark energy or dark matter, with profound consequences for fundamental cosmology \cite{Vagnozzi/2023,Dainotti/2021,Dainotti/2025}.

In this regard, it becomes essential to adopt new and independent approaches for estimating the Hubble constant. Motivated by this need, we employ a Gaussian process (GP), a fully model-independent method that relies solely on observational data and avoids assumptions about the underlying cosmological model. The freedom in this approach comes in the chosen covariance function, which determines how smooth the process is and how nearby points are correlated. 
% a new approach is very important to estimate the $H_0$ values, and for this, we use a Gaussian process, a totally model-independent method. 
A GP is a fully Bayesian approach that describes a distribution over functions and is a generalization of Gaussian distributions to function space \cite{Seikel/2012}. It is a powerful nonlinear interpolating tool without assuming a model or parameterization and is widely used in cosmology literature \cite{Seikel/2012,Seikel/2013}. Within this framework, the late-time expansion history of the universe has been reconstructed and extrapolated to infer the Hubble constant $H_0$ using BAO and SNe Ia data \cite{Bernal/2016}. However, this reconstruction depends on the sound horizon at the baryon drag epoch, $r_d$, which enters as an additional parameter. Alternative approaches have been proposed to constrain $H_0$ using only direct observational data and the distance duality relation, including SNe Ia, BAO, and cosmic chronometer (CC) measurements \cite{Renzi/2023}. Using GPs and weighted polynomial regression methods, constraints on $H_0$ have also been obtained from CC and SNe Ia data, yielding values consistent with low-$H_0$ determinations and in tension with local HST measurements \cite{Gomez-Valent/2018}. More recently, constraints on $H_0$ have been derived using DESI BAO measurements combined with unanchored SNe Ia distances and CC data, thereby avoiding assumptions related to the sound horizon scale or supernova absolute magnitude calibration \cite{Guo/2025}. Additional related studies can be found in Refs. \cite{Li/2021,Busti/2014,Jiang/2024,Cozzumbo/2025}.
% In Ref. \cite{Renzi/2023} has been proposed a method to constrain $H_0$ based solely on the distance duality relation and direct observational data, including SNe Ia, BAO, and cosmic chronometer (CC) measurements. In ref. \cite{Gomez-Valent/2018} has been constrained $H_0$ using CC and SNe Ia data with GPs and a weighted polynomial regression method, obtaining values consistent with low-$H_0$ determinations and in tension with local HST measurements. More recently, in ref. \cite{Guo/2025} has constrained the $H_0$ using DESI BAO measurements combined with unanchored SNe Ia distances and CC data, thereby avoiding assumptions related to the sound horizon scale or supernova absolute calibration. Some other work has also been done \cite{Li/2021,Busti/2014}.

In this work, we present a model-independent determination of the Hubble constant that does not rely solely on electromagnetic observations; instead, we include gravitational-wave standard siren data from the third Gravitational-Wave Transient Catalog (GWTC-3) released by the LIGO Scientific, Virgo, and KAGRA (LVK) Collaboration \cite{Abbott/2023}. The GWTC-3 catalog includes 35 new gravitational-wave candidates from the O3b observing run with an inferred astrophysical probability \( p_{\mathrm{astro}} > 0.5 \), bringing the total number of confident events to 90.
% In particular, GWTC-3 adds 35 new gravitational-wave candidates from O3b with an inferred probability of astrophysical compact binary coalescence origin exceeding \( p_{\mathrm{astro}} > 0.5 \). Of these events, 18 were previously released as low-latency public alerts, while 17 are reported for the first time in this catalog. Consequently, GWTC-3 contains a total of 90 gravitational-wave candidates. 
Gravitational-wave standard sirens provide an absolute measurement of luminosity distance and therefore offer an independent probe of the cosmic expansion history, free from the traditional cosmic distance ladder. This inclusion of the gravitational-wave observations with electromagnetic datasets improves the results as compared to previous studies and may provide new clues to fix the Hubble tension. This approach allows us to quantify the contribution of gravitational-wave standard sirens to the determination of $H_0$, assess their consistency with electromagnetic measurements, and explore their potential role in alleviating the Hubble tension.

This paper is organized as follows. In Sec. \ref{section 2}, we describe the luminosity distance reconstruction methodology adopted in this work and present a model-independent determination of the Hubble constant. The gravitational-wave standard siren data and the electromagnetic observational datasets employed in the analysis are summarized in Sec. \ref{section 3}. The results and discussion are presented in Sec. \ref{section 4}. Finally, Sec. \ref{section 5} summarizes the conclusions and outlines future research directions.
Appendix \ref{B} provides details of the Gaussian Process covariance functions and correlation coefficients used in Sec. \ref{section 2}, while additional supplementary material is given in Appendix \ref{add}.

\section{Methodology}\label{section 2}
The flat, homogeneous, and isotropic universe described by the Friedmann-Lema\^{i}tre-Robertson-Walker (FLRW) metric with line element,
\begin{equation}
    ds^2=-c^2dt^2+a^2(t)\left(dx^2+dy^2+dz^2\right),
\end{equation}
where $c$ is the speed of light and $a(t)$ is the scalar factor. The Hubble parameter is defined as $H=\frac{\dot{a}}{a}$.

For a flat FLRW cosmology, the comoving distance $D_C(z)$ is described by
\begin{equation}
    D_C(z)=c\, \int^{z}_{0}\frac{dz'}{H(z')}.
\end{equation}
The luminosity distance is directly related to the comoving distance and encodes the integrated expansion history of the universe. Photons emitted at redshift $z$ propagate through an expanding spacetime, with their trajectories governed by the inverse of the Hubble parameter. As a result, the luminosity distance accumulates the expansion information along the line of sight, making it a fundamental observable for probing late-time cosmic acceleration. The corresponding relation is given by
\begin{equation}
D_L(z)=(1+z)D_C(z)=c\,(1+z)\int_{0}^{z}\frac{dz'}{H(z')}.
\end{equation}
By differentiating this expression with respect to redshift $z$, we obtain a direct relation between the derivative of luminosity distance and Hubble parameter, given by
\begin{equation}
    \frac{d}{dz}\left(\frac{D_L(z)}{1+z}\right)=\frac{c}{H(z)}.
\end{equation}
 The luminosity distance and its derivative with respect to redshift $z$ evaluated at \(z=0\) directly isolates the Hubble constant, providing a connection between local distance measurements and the current expansion rate. This leads to the relation  
\begin{equation}\label{4}
    H_0=c\left[\left.\frac{d}{dz}\left(\frac{D_L(z)}{1+z}\right)\right|_{z=0}\right]^{-1}.
\end{equation}
Therefore, an accurate determination of $H_0$ requires knowledge of both the luminosity distance function and its first derivative at the present epoch. A common approach is to assume a specific parametric form for $D_L$ (e.g., the $\Lambda$CDM, cosmography with truncated series, or specific dark-energy models) and then evaluate its derivative at $z=0$. However, this procedure introduces model-dependent assumptions that are not entirely dictated by the data and may influence the inferred value of $H_0$.

In contrast, the GP framework provides a non-parametric and model-independent alternative, enabling the reconstruction of a smooth and continuous function directly from observational data, along with its derivatives, without assuming any predefined functional form. The GP approach relies solely on the statistical properties of the data, thereby minimizing theoretical bias in the estimation of cosmological parameters. Therefore, we adopt the \verb|GaPP| Python code\footnote{http://www.acgc.uct.ac.za/seikel/GAPP/index.html} developed by Seikel et al. \cite{Seikel/2012} in our work. In this approach, it is assumed that the values of the reconstructed function are evaluated at two different points \( z \) and \( z' \) are correlated through a covariance function \( k(z,z') \), which can be written for the luminosity distance function as
\begin{equation}
    D_L(z)\sim\mathcal{GP}\bigl(m(z),\,k(z,z')\bigr),
\end{equation}
where \( m(z) = \mathbb{E}[D_L(z)] \) denotes the prior mean function, which is often chosen to be a constant or a low-order polynomial. In our analysis, we explore the squared exponential kernel as well as Mat\'ern kernels with smoothness parameters 
$\nu = 5/2$, $7/2$, and $9/2$. The Mat\'ern kernels yield oscillatory derivative reconstructions and unstable $H_0$ estimates, making them unsuitable for the present analysis. A detailed comparison is provided 
in Appendix~\ref{kernels}. In contrast, the squared exponential kernel provides the most stable reconstruction with minimal boundary effects and well-behaved derivatives. We therefore adopt the squared exponential covariance function of the form,
\begin{equation}
k(z,z') = \sigma_f^2\exp\!\left[-\frac{(z - z')^2}{2\ell^2}\right],
\end{equation}
which characterizes the smoothness of the reconstructed function through the correlation length scale \( \ell \) and the signal variance \( \sigma_f^2 \). 

Furthermore, derivatives of $D_L(z)$ are obtained by using the same data and differentiating the same kernel, which is given by,
\begin{equation}
  D_L'(z)=  \frac{dD_L(z)}{dz}\sim\mathcal{GP}\left(m(z),\,\frac{\partial^2k(z,z')}{\partial z \partial z'}\right),
\end{equation}
and the \verb|GaPP| handles this analytically. The obtained derivative from GP is smooth, physically consistent, and includes correlation \cite{Seikel/2013}. The hyperparameters of the kernel, namely the correlation length scale \( \ell \), which indicates the distance over which significant changes occur in the function, and the signal amplitude \( \sigma_f^2 \), which characterizes the typical variation in the observed data, are determined by maximizing the marginal likelihood of the data. The optimized hyperparameters are then fixed and used in the GP reconstruction and its derivative, ensuring that the inferred luminosity distance and its derivative are described by the same underlying stochastic process.

% More importantly, at low redshifts, distance measurements are affected by the peculiar motions of galaxies, which introduce an additional source of uncertainty beyond the observational error. To account for this effect, we include a peculiar-velocity contribution for sources at low redshifts $z<0.05$, which is propagated into the luminosity-distance uncertainty. Further details on the inclusion of peculiar-velocity effects are presented in Appendix \eqref{A}.  This treatment provides a conservative and physically motivated estimate of low-redshift uncertainties, ensuring that low-redshift data points do not artificially dominate the $H_0$ inference.
More importantly, at low redshifts, distance measurements are affected by the peculiar motions of galaxies, which introduce an additional uncertainty beyond the observational error. Therefore, we include the peculiar velocity error at low redshifts $z<0.05$, which contribution is propagated into the luminosity distance uncertainty and added in quadrature to the measurement error according to
\begin{equation*}
\sigma_{D_L}^{\text{tot}}=\sqrt{\sigma_{D_L}^{2}+\left(\frac{D_L\,\sigma_{\text{pec}}}{c}\right)^2}.   
\end{equation*}
Here, $\sigma_{pec}$ denotes the characteristic peculiar velocity dispersion, for which we adopt a fiducial value of $\sigma_{\text{pec}}=300$ km s$^{-1}$ \cite{Kessler/2009}. This prescription provides a conservative and physically motivated estimate of low-redshift uncertainties, ensuring that low-redshift data points do not artificially dominate the $H_0$ inference.

 Furthermore, we compute the covariance between the reconstructed function $D_L(z_i)$ and its derivative $D'_L(z_i)$ at points $z_i$ $\forall\,i$. Further details on covariance computation are presented in Appendix \eqref{B}. We include this correlation in the error propagation of composite quantities, which is defined as
\begin{equation}
   \sigma_{r'(z)}^2=\frac{\sigma^2_{D'_L}}{(1+z)^2}+\frac{\sigma^2_{D_L}}{(1+z)^4}-\frac{2}{(1+z)^3}\sigma_{D_LD_L'},
\end{equation}
where $r(z)=D_L/(1+z)$ and $\sigma_{D_LD_L'}$ represents the covariance between $D_L$ and $D_L'$, ensuring a statistically consistent treatment of correlated uncertainties.
%$\frac{\mathrm{d}^m}{\mathrm{d}x^m}\frac{\mathrm{d}^n}{\mathrm{d}x'^n}k(z,z')$ retains Gaussian form, allowing direct reconstruction n$^{th}$ order derivative of $f(z)$ with uncertainty estimates . 
% This framework enables a robust, data-driven reconstruction of cosmological quantities, making Gaussian processes a powerful tool for probing the expansion history of the universe. This work is not only a GP with supernovae data, which was previously done in many works. 

In this work, we implement this framework to estimate a model-independent measurement of the Hubble constant $H_0$ 
%by reconstructing the luminosity distance relation $D_L(z)$ and its derivative 
via GP trained on a combined dataset of gravitational-wave standard sirens and electromagnetic (EM) probes. 
%The GP framework enables a direct reconstruction of the luminosity distance from observations without assuming a specific cosmological model. Using the reconstructed $D_L(z)$ and its associated uncertainties, we then infer the corresponding expansion history and extract the present-day value of the Hubble constant. 
This data-driven approach provides an independent determination of $H_0$ and offers a new way to investigate the origin of the Hubble tension.

\section{Data}\label{section 3}
\subsection{Gravitational Wave Data}

The first direct detection of gravitational waves from a binary black hole (BBH) coalescence in 2015 \cite{Abbott/2016} marked the advent of gravitational-wave astronomy. Since then, the Advanced Laser Interferometer Gravitational-Wave Observatory (Advanced LIGO) \cite{Aasi/2015} and the Advanced Virgo detectors \cite{Acernese/2015} have opened a new observational window on the Universe \cite{Abbott/2016a,Abbott/2017,Abbott/2017a,Abbott/2017b,Abbott/2019}. Observations of compact binary mergers enable stringent tests of gravity in the strong-field regime \cite{Abbott/2018,Abbott/2019a} and provide key insights into the formation channels, merger rates, and population properties of BBH systems \cite{Abbott/2019b}. In addition to BBH mergers, Advanced LIGO and Advanced Virgo detected the first gravitational-wave signal from a binary neutron star (BNS) coalescence, GW170817 \cite{Abbott/2017c}, which also constituted the first joint detection of gravitational waves and electromagnetic emission. This milestone event firmly established multi-messenger astronomy and demonstrated the profound impact of gravitational-wave observations on fundamental physics, astrophysics, and cosmology. Moreover, the public release of LIGO and Virgo data \cite{Abbott/2021,Trovato/2020} has facilitated independent analyses by the broader scientific community, leading to additional searches for gravitational-wave signals and the identification of new candidate events.

In this work, we utilize data from the third Gravitational-Wave Transient Catalog (GWTC-3) released by the LIGO Scientific, Virgo, and KAGRA (LVK) Collaboration \cite{Abbott/2023}. GWTC-3 reports transient gravitational-wave signals detected up to the end of the third observing run (O3) of the LIGO--Virgo network and represents a significant update to the earlier GWTC-2 \cite{Abbott/2021a} and GWTC-2.1 \cite{Abbott/2021b} catalogs by including events identified during the second part of O3 (O3b). In particular, GWTC-3 adds 35 new gravitational-wave candidates from O3b with an inferred probability of astrophysical compact binary coalescence origin exceeding \( p_{\mathrm{astro}} > 0.5 \). Of these events, 18 were previously released as low-latency public alerts, while 17 are reported for the first time in this catalog. Consequently, GWTC-3 contains a total of 90 gravitational-wave candidates with \( p_{\mathrm{astro}} > 0.5 \).

From a cosmological perspective, compact binary coalescences act as \emph{standard sirens}, as their gravitational-wave signals provide an absolute measurement of the luminosity distance that is independent of the cosmic distance ladder. When combined with redshift information obtained from electromagnetic counterparts or host galaxy associations, these distance measurements enable direct constraints on the cosmic expansion rate. In this study, we incorporate the GWTC-3 standard-siren data into our Gaussian Process reconstruction framework to infer the luminosity--distance relation in a model-independent manner. However, the current GW sample is limited to relatively low redshifts ($z \lesssim 1.2$, with a median of $z \approx 0.30$), which can lead to large uncertainties when used alone. To obtain a stable and well-constrained reconstruction over a wider redshift range, we therefore complement the GW data with electromagnetic observations, which are discussed below. This combined approach takes advantage of the complementary nature of GW and EM probes, improving the robustness of the reconstruction and enabling a more reliable determination of the Hubble constant.

The luminosity distance posteriors used in this work are taken directly from the publicly available GWTC-3 catalog, as reported by the LVK Collaboration. The standard LVK parameter estimation pipeline assumes a flat-in-log prior on source-frame component masses, which introduces a redshift-dependent prior on luminosity distance. While this does not affect the model-independence of the GP reconstruction itself, it represents a subtle prior dependence embedded in the input GW data. A fully prior-independent treatment would require reweighting the posterior samples against a cosmology-agnostic reference prior, which we identify as an important direction for future work. We note, however, that this effect is expected to be subdominant relative to the large statistical uncertainties currently associated with GW standard sirens, and is unlikely to qualitatively alter our conclusions.
\subsection{Electromagnetic Data }
To improve the stability of the reconstruction across a wider redshift range, we complement the siren dataset with:
\paragraph{Cosmic Chronometer (CC):} CC measurements provide direct estimates of the Hubble parameter $H(z)$ by using the differential ages of galaxies, which are determined by the evolution of passively aging star populations and their spectroscopic redshifts \cite{CC1,CC2,CC3,CC4,CC5,CC6}. Current CC datasets span a redshift range from the local universe up to redshifts of about $z\approx 2$. The relation between the Hubble function and the luminosity distance is defined as
    \begin{equation}
        D^{CC}_L(z)= c\,(1+z) \int^{z}_{0}\frac{dz'}{H(z')}.
    \end{equation}
    Using this relation, Hubble measurements can be converted into luminosity distances by numerically evaluating the integral, with associated uncertainties obtained by integrating the error of the $H(z)$ functions. 
\paragraph{DESI BAO DR2:}  We use BAO measurements from the second data release of DESI, which includes observations of galaxies and quasars \cite{Karim/2025}, as well as Lyman-$\alpha$ tracers \cite{Karim/2025b}. These measurements cover both isotropic and anisotropic BAO constraints over $0.295 \leq z \leq 2.330$, divided into nine redshift bins \cite{Karim/2025}.
The DESI BAO DR2 measurements provide constraints that are expressed in terms of the transverse comoving distance $D_M/r_d$, the Hubble horizon $D_H/r_d$, and the angle-averaged distance $D_V(z)/r_d$, all normalized to the comoving sound horizon at the drag epoch, $r_d$. 
%DESI BAO measurements provide constraints on the transverse comoving distance through observations of the quantity $D_M(z)/r_d$, where $D_M(z)$ denotes the comoving angular diameter distance and $r_d$ is the sound horizon at the drag epoch \cite{Karim/2025}. 
We could convert these measurements into luminosity distances using the transverse comoving distance $D_M/r_d$ quantity, given by 
\begin{align}
    D^{\rm DESI}_L(z)&=(1+z)D_M(z)\\
    &= (1+z)r_d\left(\frac{D_M(z)}{r_d}\right)\nonumber,
\end{align}
 where the sound horizon $r_d$ is treated as a nuisance parameter and marginalized over to preserve model independence. The associated uncertainties are obtained by calculating the error of $D_M$ as 
 \begin{equation}
     \sigma_{D^{\rm DESI}_L}=(1+z)r_d\left(\frac{\sigma_{D_M}}{r_d}\right).
 \end{equation}
\paragraph{Type Ia Supernovae (SNe Ia):} We use three compilations of SNe Ia, namely the Pantheon+SH0ES, Union 3.0, and DES-SN5YR samples, to determine how each of these catalogs affects the $H_0$ measurement when combined with GW data.
%all of which provide measurements of the distance modulus $\mu(z)$ over a wide redshift range.
The distance modulus derived from the Pantheon+SH0ES data set is based on 1701 light curve measurements from 1550 distinct supernovae, covering a redshift range of $z\in[0.001, 2.2613]$ \cite{Scolnic/2018}. A notable aspect of this data set is the use of SH0ES Cepheid host distance anchors \cite{Riess/2022} to calibrate the absolute magnitude of SNe Ia, rather than imposing a prior on the $H_0$ value from SH0ES. This approach allows for a direct determination of the Hubble constant independent of early-Universe assumptions, thereby making Pantheon+SH0ES particularly relevant for investigations of the Hubble tension.
% Additionally, the host distance of Cepheid variable stars was determined from 77 data points associated with supernovae in their host galaxies, with a redshift range of $0.00122 < z < 0.01682$. 
Regarding the Union 3.0 SN dataset, it represents a homogeneous and high-quality compilation of SNe Ia observations constructed by the Supernova Cosmology Project \cite{Rubin/2023}. It consists of 22 binned data points derived from a total of 2087 SNe Ia spanning the redshift range $0.05 \leq z \leq 2.26$, with 1360 supernovae overlapping with the Pantheon+ sample. Union 3.0 shares a larger fraction of the data with Pantheon+ but uses a different approach based on the Bayesian hierarchical modeling framework Unity1.5 \cite{Rubin/2015}. This dataset is referred to as Union3.
% Regarding the Union 3.0 SN dataset, it represents a homogeneous and high-quality compilation of Type Ia supernova observations constructed by the Supernova Cosmology Project \cite{Rubin/2023}. This dataset builds upon the earlier Union and Union2 compilations by incorporating improved photometric calibration procedures, refined treatments of systematic uncertainties, and an expanded, uniformly processed SN sample. The Union3 catalog consists of more than 1400 spectroscopically confirmed SN Ia, spanning the redshift interval $z \in [0.01,\,1.4]$. These datasets offer complementary coverage and serve as a consistency check on the luminosity distance reconstruction. 
Finally, the DES-SN5YR sample, released as part of the Dark Energy Survey (DES) Year 5 data, includes a homogeneously selected sample of 1635 photometrically classified SNe Ia with redshifts in the range $0.1 < z < 1.3$ \cite{DESY5}. This sample is supplemented by 194 low-redshift SNe Ia, shared with the Pantheon+ sample, spanning $0.025 < z < 0.1$. This combined dataset is referred to as DESY5. Although the underlying data differ substantially, the Pantheon$+$ and DESY5 analysis methodologies are quite similar in nature. However, DESY5 employs the SALT3 \cite{SALT3} light-curve fitting model, whereas Pantheon+ utilizes SALT2 \cite{SALT2}, among other methodological distinctions.

The luminosity distance of SNe Ia is related to the distance modulus given by
    \begin{equation}
        \mu_{\rm{SN}}(z)=25+5\,\log_{10}\left(\frac{D^{\text{SN}}_L(z)}{1 Mpc}\right).
    \end{equation}
    Using this relation, the SNe distance modulus measurements are converted into luminosity distances
    \begin{equation}
        D^{\text{SN}}_L(z)=10^{(\mu_{\text{SN}}-25)/5},
    \end{equation}
    with the corresponding uncertainties obtained through standard error propagation of the observational errors associated with the distance modulus as
    \begin{equation}
        \sigma_{D^{\text{SN}}_L}=\frac{10^{(\mu_{\text{SN}}-25)/5}\,\log10}{5}\,\sigma_{\mu_{\text{SN}}}.
    \end{equation}
\paragraph{Gamma-Ray Bursts (GRB):} In addition, we employ GRB observations as high-redshift distance indicators, which substantially extend the cosmological leverage beyond the supernova regime. GRBs are standardized using empirical luminosity correlations, most notably the Amati relation between the rest-frame spectral peak energy $E_{p,i}$ and the isotropic-equivalent radiated energy $E_{\rm iso}$ \cite{Amati/2002,Amati/2006}. After calibrating this correlation in a cosmology-independent manner using low-redshift SNe~Ia, the isotropic energy is used to infer the luminosity distance according to
\begin{equation}
D^{\mathrm{GRB}}_L(z) = \left( \frac{E_{\rm iso}(1+z)}{4\pi S_{\rm bolo}} \right)^{1/2},
\end{equation}
where $S_{\rm bolo}$ denotes the observed bolometric fluence. The corresponding distance modulus is then obtained through the standard relation $\mu(z)=25+5\log_{10}D_L(z)$. The uncertainties in the GRB luminosity distances are estimated by propagating the measurement errors in $E_{p,i}$ and $S_{\rm bolo}$, together with the intrinsic scatter of the calibration relation. Owing to their detectability up to very high redshifts ($z \gtrsim 6$), GRBs provide a complementary probe of the cosmic expansion history, bridging the redshift gap between SNe Ia and other late-time cosmological observables \cite{Demianski/2017a,Demianski/2017b}.

Throughout this work, the luminosity distance $D_L(z)$ derived from all observational datasets is consistently expressed in units of megaparsecs (Mpc).

\section{Results and Discussions}\label{section 4}
Our approach is to determine the Hubble constant $H_0$ in a model-independent manner directly from observational data and is, in principle, not affected by local systematics. Unlike the traditional $\Lambda$CDM model or distance-ladder methods, which rely on specific cosmological assumptions or a series of interconnected measurements, our method avoids such dependencies. We have used the most promising GP method, which can reconstruct a function and its derivatives with high accuracy (see Appendix \ref{add}), which indicates that the GP method is a promising method in cosmological research. Most importantly, inspired by previous work where high-redshift and direct luminosity distance observations were found to be crucial for constraining the Hubble constant, we have jointly exploited the GW observation by the third-generation (GWTC3) and electromagnetic observations.
GWs offer an absolute, self-calibrated measurement of luminosity distances that does not rely on the cosmic distance ladder or assumptions regarding the cosmological model. Nevertheless, present GW observations are constrained by statistical uncertainties and limited redshift coverage. When combined with EM observations, we capitalize on their double strengths: the absolute distance scale from standard sirens and the high-precision redshift information from EM observations. This highlights the importance and timeliness of our joint GW-EM strategy in addressing these challenges.

\begin{figure*}
    \centering
    \includegraphics[width=0.8\linewidth]{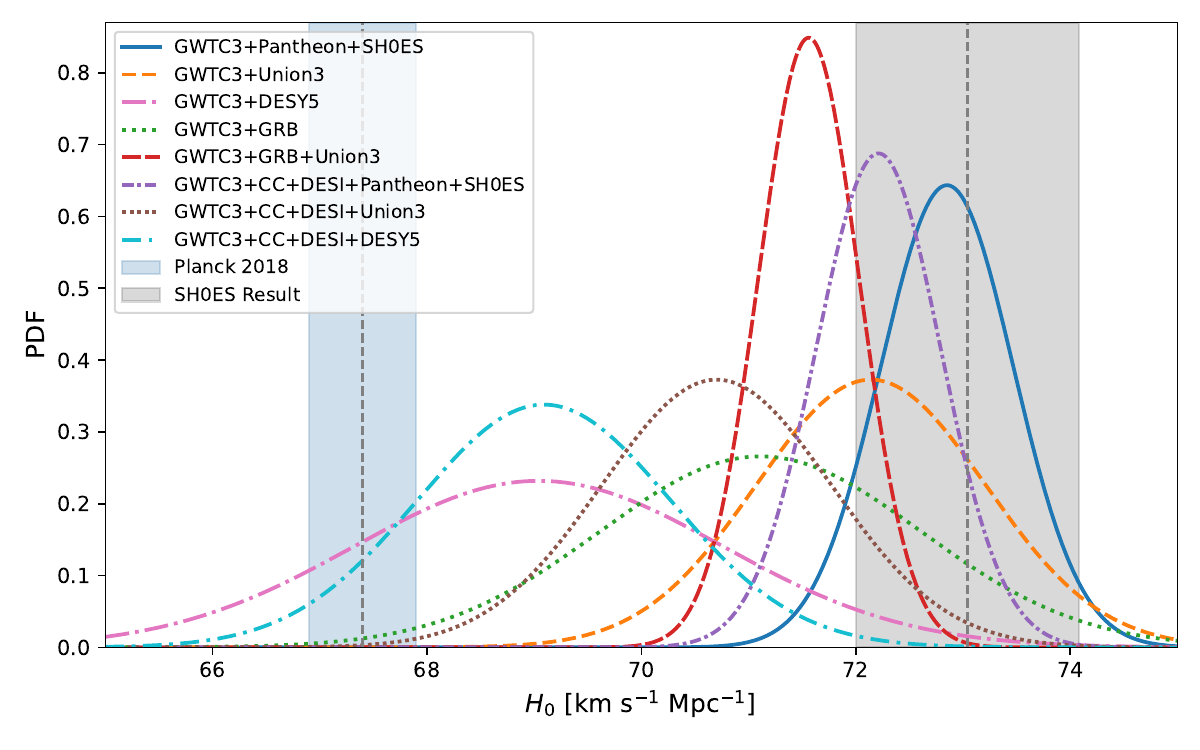}
    \caption{\justifying Posterior distributions of the Hubble constant \(H_0\). 
The black dotted line and shaded band indicate the SH0ES constraint and its \(1\sigma\) uncertainty.} \label{fig1}
\end{figure*}

First, we have reconstructed the luminosity distance $D_{L}(z)$ and its derivative with respect to redshift $z$ by using the different combinations of GW and EM data\footnote{When considering only the GWTC3 catalog, the resulting value of the Hubble constant is $H_0=68.16\pm11.08$. The uncertainty associated with GW catalogs remains substantial, primarily due to the lack of electromagnetic counterparts for most events. This absence necessitates indirect redshift inference through galaxy catalogs, which, together with significant luminosity-distance uncertainties and a limited number of low-redshift sources, increases the overall uncertainty. Incorporating EM data into GW analyses enhances the measurement of the Hubble constant by enabling direct host-galaxy redshift determination. Therefore, the combined use of GW and EM observations provides a more robust and transparent estimate of $H_0$.}. Due to the combination of two different measurements, it may introduce an additional uncertainty beyond the observational error at low redshifts. Therefore, we add total uncertainty, which includes both observational errors and an additional contribution from peculiar velocities at low redshift $z\leq 0.05$. This treatment, similar to the error analysis commonly utilized in distance ladder approaches such as Cepheid or TRGB calibrations, provides a robust framework that anchors our hybrid dataset within a familiar context. 
%By drawing a parallel to these well-trusted calibration techniques, our method positions itself as a coherent next rung on the distance ladder, rather than an isolated technique. 
This is particularly important near $z\approx 0$, where the inference of $H_0$ is most sensitive.
%First, we have combined the GW data and SNe Ia compilations\footnote{Importantly, we only include one of these SNe Ia compilations at a time, as they share overlapping supernovae and correlated systematics. We perform separate analyses using Pantheon+SH0ES or Union3 or DESY5, each combined with GWTC-3 standard sirens.}. 
Further, the GP yields the posterior mean functions along with their full covariance matrix, allowing the statistical correlation between $D_L(z)$ and $dD_L/dz$ to be explicitly taken into account. This covariance plays a crucial role in ensuring a consistent propagation of uncertainties into the estimated Hubble constant $H_0$. By evaluating the kinematical relation between the reconstructed $D_L(z)$ and $dD_L/dz$ at $z=0$ (see in Eq. \eqref{4}), we estimate $H_0$ value. To statistically sample $H_0$, random realizations of $D_L(z)$ and $dD_L/dz$ are drawn from their GP-predicted Gaussian distributions at the lowest redshift point. This procedure effectively generates $5000\times5000$ possible realizations of $H_0$, fully capturing the propagated uncertainties and correlations encoded in the reconstruction. From this ensemble, a one-dimensional posterior probability distribution function (PDF) for the Hubble constant is constructed. The
PDFs of the Hubble constant with all combined results are depicted in Figure \ref{fig1}. The mean of this posterior is taken as the best-fit value of $H_0$, while its standard deviation provides the corresponding $1\sigma$ uncertainty, which is explicitly mentioned in Table \ref{Table 1}. 

Our obtained Hubble constant $H_0$ values lie in between the Planck and SH0ES measurements, with a clear preference toward the SH0ES result, $H_0=73.04\pm1.04$ km s$^{-1}$ Mpc$^{-1}$. For most of data combinations, the tension between our obtained values of $H_0$ and the SH0ES result is less than $2\sigma$, indicating that our measurement is strongly consistent with the SH0ES determination, as summarized in Table~\ref{Table 1}. An exception is the GWTC3+CC+DESI+DESY5 combination, for which the obtained $H_0$ slightly favors the Planck measurement. More importantly, our analysis significantly reduces the uncertainty of $H_0$ without imposing any prior as compared to several previous model-independent studies based solely on EM data \cite{Gomez-Valent/2018,Guo/2025,Li/2021,Busti/2014}. To explicitly quantify the contribution of the GW data, we run GP for more general Pantheon+SH0ES+CC+DESI combination without GW data, and we obtain the $H_0 = 72.8 \pm 1.02$ km s$^{-1}$ Mpc$^{-1}$, comparing this with the corresponding GW+EM combination, GWTC3+CC+DESI+Pantheon+SH0ES, which yields $H_0 = 72.21 \pm 0.58$ km s$^{-1}$ Mpc$^{-1}$, we find that the inclusion of GWTC-3 data reduces the uncertainty from $1.02$ to $0.58$ km s$^{-1}$ Mpc$^{-1}$, representing a reduction of approximately $15-20\%$ in the $1\sigma$ uncertainty in an average.  Hence, the inclusion of gravitational-wave standard sirens data from GWTC3 in our analysis provides an absolute calibration of cosmic distances, leading to a noticeably tighter constraint on $H_0$. This result highlights the growing role of GW observations and suggests that their combination with electromagnetic data can play a meaningful role in addressing the current Hubble tension.

We note an important caveat regarding the GW data: for BBH events without electromagnetic counterparts, redshift inference relies on host galaxy catalog associations or on population-level assumptions about the intrinsic BBH mass distribution. The assumed BBH population model, including the shape of the mass function and the redshift evolution of the merger rate, can be degenerate with cosmological parameters, and an incorrect population prior may introduce systematic biases in the inferred luminosity distances. A rigorous treatment of this systematic would require a simultaneous population-and-cosmology inference, which is beyond the scope of the present work. We identify this as an important caveat and a direction for future investigation, particularly as next-generation GW detectors provide larger event catalogs with tighter constraints on the BBH population.
%\begin{widetext}

\begin{table*}[t]
\begin{center}
  \caption{\justifying Estimated values of the Hubble constant $H_0$ obtained from different combinations of cosmological datasets. The quoted tensions are computed with respect to the SH0ES result.}
    \label{Table 1}
    \begin{tabular}{l c c c }
\hline\hline 
Dataset      & $H_0\pm\sigma_{H_0}$ [km s$^{-1}$ Mpc$^{-1}$] & Tension vs. SH0ES ($\sigma$) & Tension vs. Planck ($\sigma$) \\[1ex] 
\hline\hline
GWTC3+Pantheon+SH0ES  & $72.85\pm0.61$ & $0.16\sigma$ & $6.9\sigma$ \\[1ex]

GWTC3+Union3 & $72.14\pm1.07$ & $0.6\sigma$ & $4.0\sigma$\\[1ex]

GWTC3+DESY5 & $69.04\pm1.72$ & $1.98\sigma$ & $0.9\sigma$ \\[1ex]

GWTC3+GRB & $71.11\pm1.5$ & $1.06\sigma$ & $2.3\sigma$ \\[1ex]

GWTC3+GRB+Union3 & $71.56\pm0.47$ & $1.3\sigma$ & $6.0\sigma$  \\[1ex]

GWTC3+CC+DESI+Pantheon+SH0ES & $72.21\pm0.58$ & $0.7\sigma$ & $6.3\sigma$ \\[1ex]

GWTC3+CC+DESI+Union3 & $70.70\pm1.07$ & $1.6\sigma$ &  $2.8\sigma$\\[1ex]

GWTC3+CC+DESI+DESY5 & $69.09\pm1.18$ & $2.5\sigma$ & $1.32\sigma$ \\[1ex]

\hline \hline
\end{tabular}
\end{center}
\end{table*}
\section{Conclusions}\label{section 5}
In this work, we have studied a model-independent method to measure the Hubble constant, thereby avoiding potential biases associated with traditional model-dependent approaches. While this study does not directly claim to resolve the Hubble tension, instead, the model-independent measurements of the $H_0$, derived from combined EM and GW data, yield robust and consistent $H_0$ estimates that strongly favor the SH0ES results. Moreover, this approach quantifies the impact of GW data on the determination of the Hubble constant, examines its consistency with EM measurements, and could give a new perspective to handle the Hubble tension issue by relying solely on observational inputs. These $H_0$ estimates can be directly applied to calibrate distance-redshift relations, constrain late-time cosmological models, and serve as informative priors in dark energy and modified gravity analyses. 

Although the current GW data are still limited, their inclusion already shows a meaningful impact and consistency with EM measurements. The reconstruction framework adopted in this work can be extended to include future GW observations from next-generation detectors, which are expected to significantly improve the precision of standard siren measurements. In addition, including larger and more precise EM datasets, such as upcoming supernova compilations and BAO measurements, will allow for more rigorous model-independent constraints on the Hubble constant. The methodology can also be generalized to test consistency among different distance indicators and to investigate potential systematics affecting late-time cosmological probes. These extensions will provide further understanding of the origin of the Hubble tension and the robustness of model-independent reconstruction techniques.

\acknowledgments
Gaurav N. Gadbail and Kazuharu Bamba acknowledge the support by the JSPS KAKENHI Grant Number 25KF0176. The work of Kazuharu Bamba was supported in part by the JSPS KAKENHI Grant Number 24KF0100 and Competitive Research Funds for Fukushima University Faculty (25RK011).

% this approach is likely to become an increasingly valuable tool for precision cosmology and for better understanding the origin of the Hubble tension.
\appendix
\section{Kernel Sensitivity Analysis}\label{kernels}

To assess the robustness of our results to the choice of GP covariance function, we repeated the $H_0$ inference using Mat\'ern kernels, 
\begin{multline*}
    k(z,z')=\sigma_f^2\,\frac{2^{1-\nu}}{\Gamma(\nu)}\left(\frac{\sqrt{2\nu(z-z')^2}}{\ell}\right)^{\nu}\\\times K_{\nu}\left(\frac{\sqrt{2\nu(z-z')^2}}{\ell}\right),
\end{multline*}
where $K_{\mu}$ is the modified Bessel function and $\nu$ is a positive smoothness parameter which defines the shape of the covariance function. \textit{The Mat\'ern covarience function is k-times differentiable iff $\nu>k$.} That means the differentiability depends on choice of smoothness parameter $\nu$. We consider smoothness parameters $\nu = 5/2$, $7/2$, and $9/2$, corresponding to sample functions that are $C^2$ (2-times), $C^3$ (3-times), and $C^4$ (4-times) differentiable, respectively, across all dataset combinations considered in this work. The results are summarized in Table~\ref{Table 2}.

Several features are immediately apparent from this comparison. First, the Mat\'ern kernels produce $H_0$ estimates that are highly sensitive to the choice of $\nu$ \cite{Eoin}, with mean values shifting by as much as 
$\sim 8$ km s$^{-1}$ Mpc$^{-1}$ across different smoothness parameters for the same dataset combination (e.g., GWTC3+Union3 varies from $64.81$ to $73.26$ km s$^{-1}$ Mpc$^{-1}$). Second, several combinations yield unrealistically large uncertainties under the Mat\'ern kernels (e.g., $\sigma_{H_0} \sim 10$--$12$ km s$^{-1}$ Mpc$^{-1}$ for GWTC3+Union3 and GWTC3+DESY5), indicating that the reconstruction has not converged to a physically meaningful solution. These instabilities 
originate from the finite differentiability of the Mat\'ern class, which introduces spurious oscillations in the reconstructed derivative $dD_L/dz$ near $z \approx 0$, directly contaminating the $H_0$ inference. In contrast, the squared exponential kernel is infinitely differentiable ($C^\infty$), yields well-behaved derivatives (see Fig.~\ref{fig2}) and produces stable, consistent $H_0$ estimates across all dataset combinations, as reported in Table~\ref{Table 1}. We therefore conclude that the squared exponential kernel is the most appropriate choice for derivative-based $H_0$ reconstruction.

\section{GP Covariances and correlation coefficient}\label{B}
In this appendix, we describe the GP covariances and the correlation coefficient between the reconstructed luminosity distance \(D_L\) and its derivative \(D'_L\).
The GP reconstructed luminosity distance \(D_L\) and its derivative \(D'_L\) are not statistically independent quantities; rather, they are intrinsically correlated through the GP covariance structure. Accounting for this correlation is essential for a statistically consistent propagation of uncertainties in derived cosmological quantities, particularly in the reconstruction of the Hubble expansion rate and the determination of the Hubble constant \(H_0\). The \verb|GaPP| code explicitly computes the covariance between these two quantities.

For a Gaussian Process object initialized as \verb|g|, the command
\verb|fcov = g.f_covariances(fclist = [0,1])|, where $0$ denotes the reconstructed function and $1$ denotes its first derivative,
computes the covariances between \(D_L(z_i)\) and \(D'_L(z_i)\) evaluated at the redshift points \(z_i\). For each \(i\), the array \verb|fcov[:, 0, 1]| contains the corresponding \(2\times2\) covariance matrix,
\begin{equation}
\begin{pmatrix}
\mathrm{Var}\!\left[D_L(z_i)\right] & \mathrm{Cov}\!\left[D_L(z_i),D'_L(z_i)\right] \\
\mathrm{Cov}\!\left[D_L(z_i),D'_L(z_i)\right] & \mathrm{Var}\!\left[D'_L(z_i)\right]
\end{pmatrix}.
\end{equation}

We define the correlation coefficient to quantify the degree of correlation between \(D_L\) and \(D'_L\) as
\begin{equation}
\rho = \frac{\mathrm{Cov}\!\left[D_L(z),D'_L(z)\right]}{\sigma_{D_L}\,\sigma_{D'_L}},
\end{equation}
which satisfies \( -1 \leq \rho \leq 1 \) by construction and provides a dimensionless measure of the strength of the correlation.

Figure~\ref{fig2} displays the resulting correlation coefficient for all data combinations considered in this work. We observe a strong and smooth positive correlation across the full redshift range, particularly at low redshifts where the data density is highest. The correlation gradually decreases toward higher redshifts due to increasing uncertainties and boundary effects inherent to the GP reconstruction, as expected.

\begin{widetext}

\begin{figure}[H]
    \centering
    \includegraphics[width=0.425\linewidth]{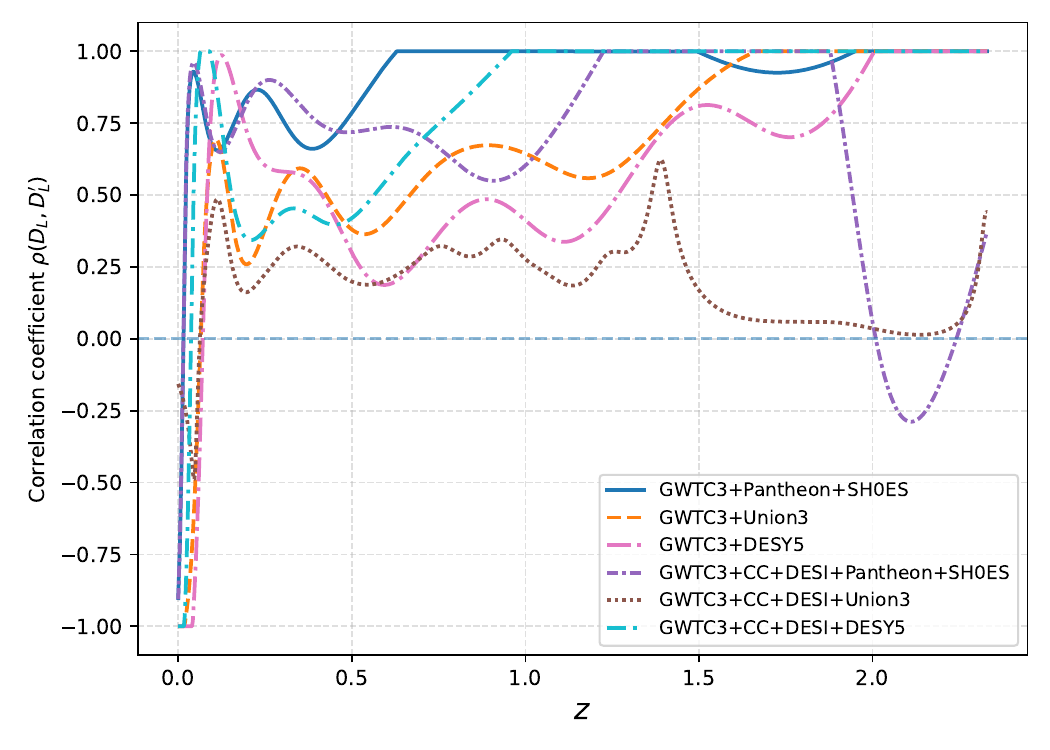}
     \includegraphics[width=0.425\linewidth]{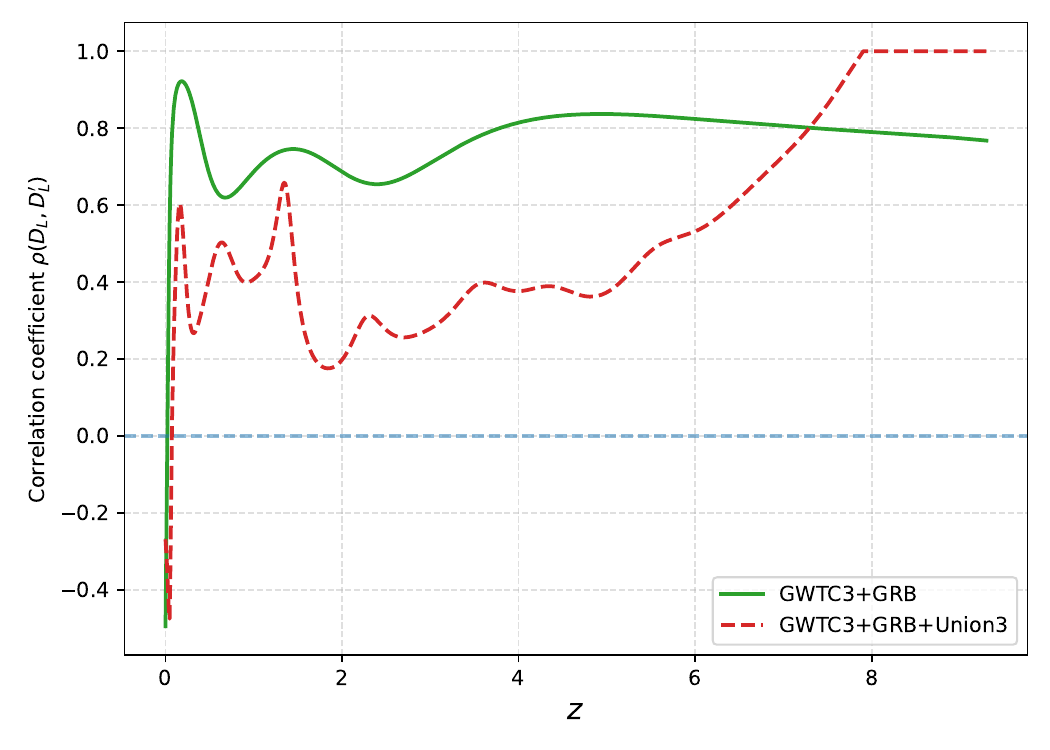}
    \caption{\justifying The correlation coefficient between the reconstructed luminosity distance $D_L$ and its derivative $D'_L$ for different data combinations. }
    \label{fig2}
\end{figure}

\begin{table}[H]
\begin{center}
\caption{\justifying Estimated values of the Hubble constant $H_0$ (in km s$^{-1}$ Mpc$^{-1}$) obtained using Mat\'ern covariance kernels with smoothness parameters $\nu = 5/2$, $7/2$, and $9/2$.}
\label{Table 2}
\begin{tabular}{l c c c}
\hline\hline 
Dataset & Mat\'ern $5/2$ & Mat\'ern $7/2$ & Mat\'ern $9/2$ \\[1ex] 
\hline\hline
GWTC3+Pantheon+SH0ES  
    & $74.25\pm 2.80$ & $72.62\pm 1.18$ & $72.42\pm 0.87$ \\[1ex]
GWTC3+Union3 
    & $73.26\pm10.52$ & $66.88\pm 4.29$ & $64.81\pm 2.42$ \\[1ex]
GWTC3+DESY5 
    & $64.17\pm12.29$ & $63.58\pm 5.35$ & $64.78\pm 3.65$ \\[1ex]
GWTC3+GRB 
    & $67.49\pm 2.90$ & $68.61\pm 1.67$ & $69.25\pm 1.62$ \\[1ex]
GWTC3+GRB+Union3 
    & $67.49\pm 3.92$ & $70.77\pm 1.24$ & $70.37\pm 0.79$ \\[1ex]
GWTC3+CC+DESI+Pantheon+SH0ES 
    & $74.32\pm 2.86$ & $72.70\pm 1.17$ & $72.44\pm 0.86$ \\[1ex]
GWTC3+CC+DESI+Union3 
    & $73.39\pm11.47$ & $66.73\pm 5.02$ & $64.17\pm 3.30$ \\[1ex]
GWTC3+CC+DESI+DESY5 
    & $63.69\pm 6.88$ & $66.28\pm 2.83$ & $68.26\pm2.02$ \\[1ex]
\hline\hline
\end{tabular}
\end{center}
\end{table}

\end{widetext}
\onecolumngrid
\section{Additional Figures}\label{add}
In this appendix, we present the figures of the reconstructed luminosity distance and its derivative to demonstrate the smoothness and physical consistency of our model-independent reconstruction. To check the physical reliability, we examine the possibility of unphysical behavior. We find that the probability of obtaining negative values of $D_L'(z)$ over the full redshift range is below $1\%$, indicating that the reconstruction is consistent with the basic physical expectation of an expanding universe.
% The physical consistency of the reconstruction is further assessed by evaluating the probability of unphysical behavior. The probability of negative values for $D_L'(z)$ across the full redshift range is found to be $P(D_L'\leq 0)<1\%$, which confirms that the GP reconstruction adheres to the fundamental physical requirements of an expanding universe.
\begin{figure}[H]
    \centering
    \includegraphics[width=0.44\linewidth]{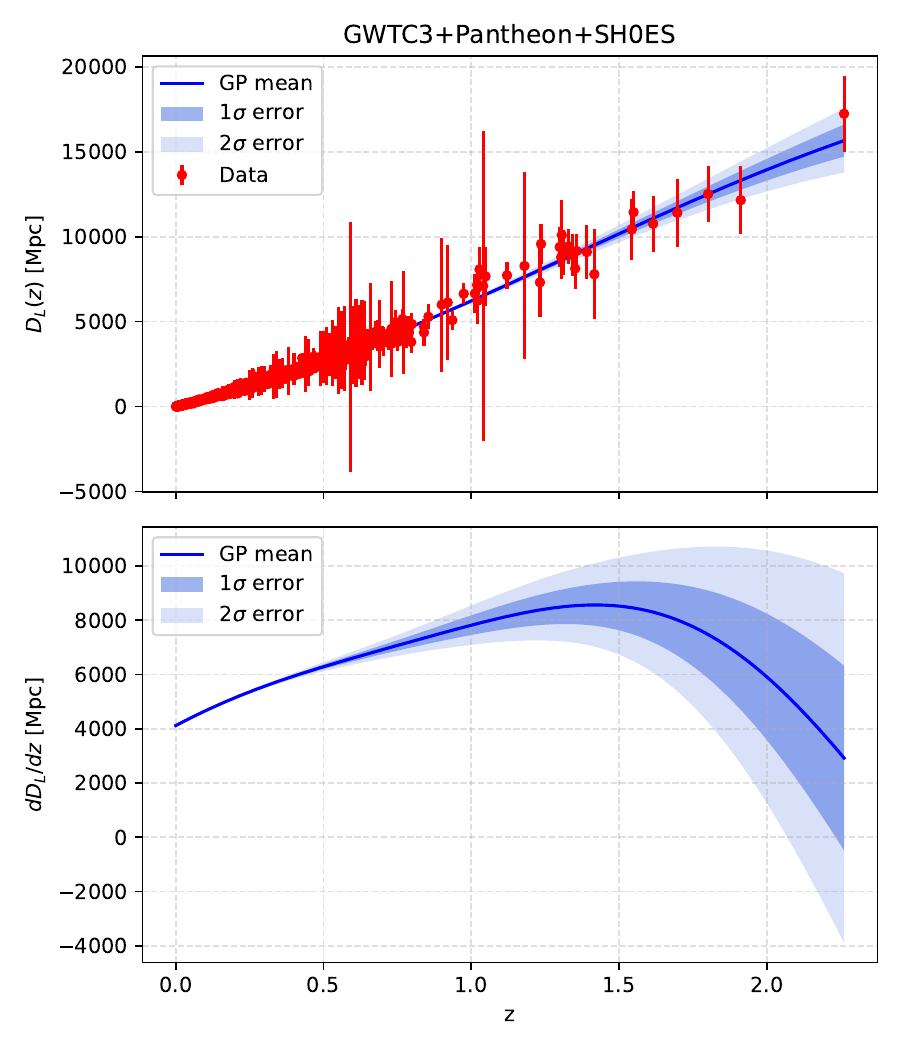}
     \includegraphics[width=0.44\linewidth]{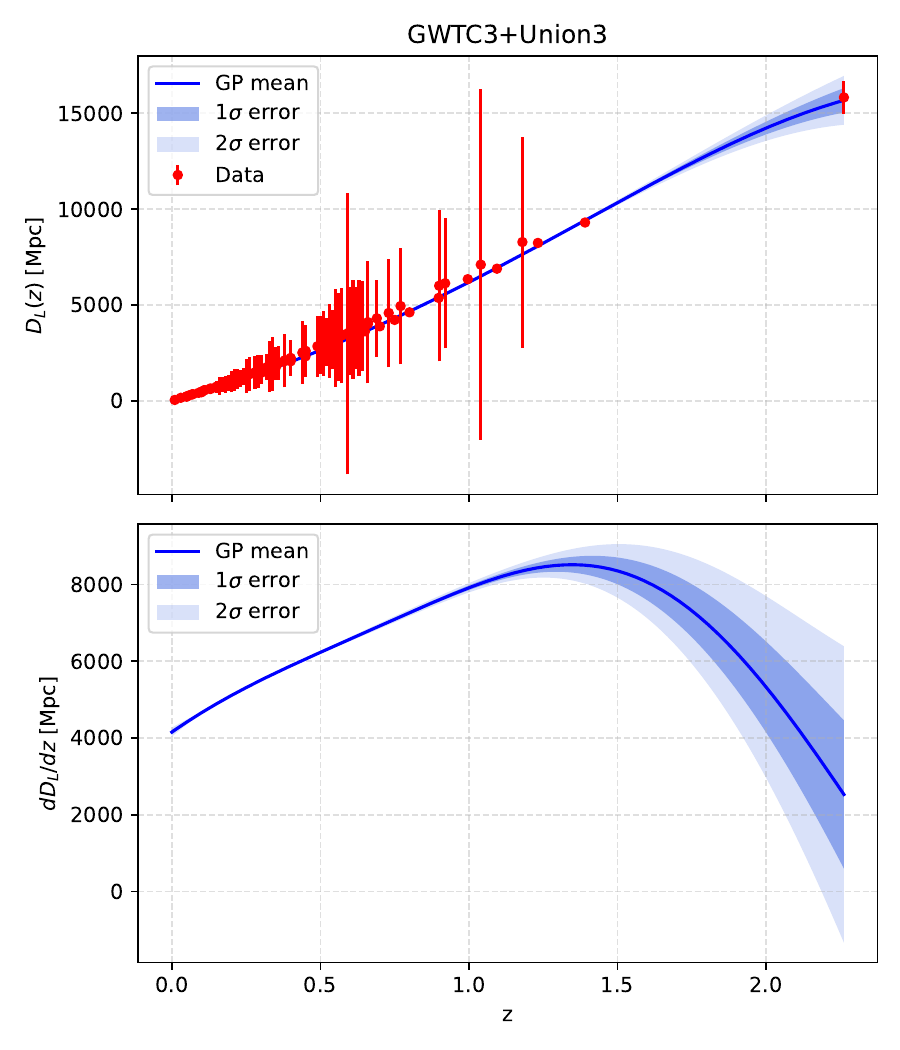}
      \includegraphics[width=0.44\linewidth]{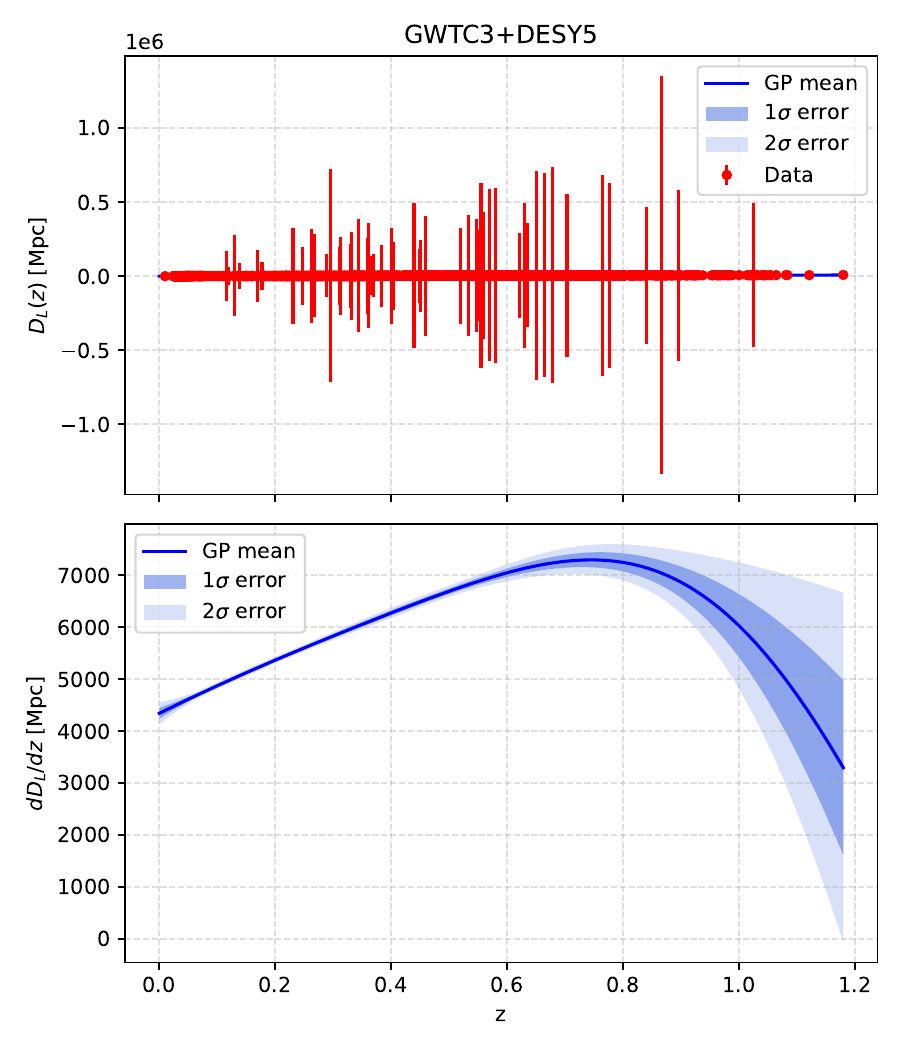}
     \includegraphics[width=0.44\linewidth]{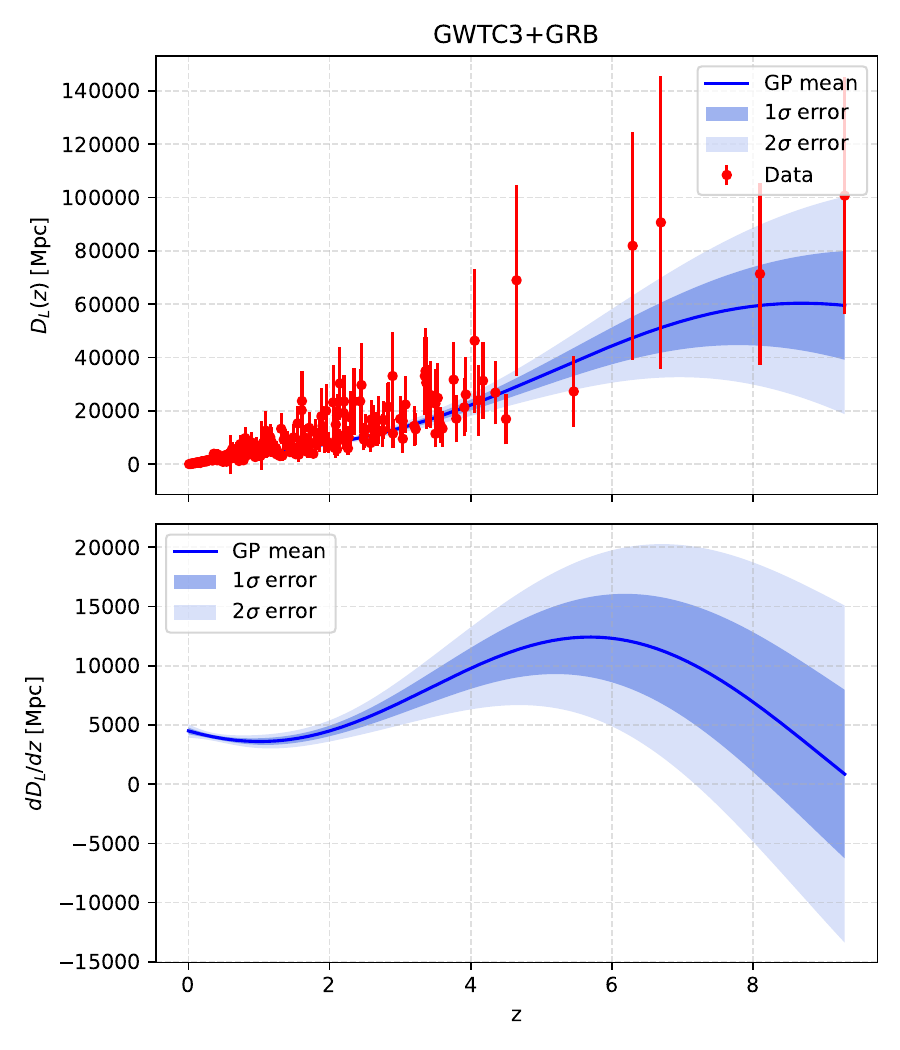}
    \caption{\justifying Reconstructed luminosity distance $D_L$ and its derivative $D_L'$. The blue line represents the mean reconstruction, while the shaded regions correspond to the $1\sigma$ and $2\sigma$ confidence intervals.   }

    \label{fig2}
\end{figure}
% \section*{Data Availability}
%  There are no new data associated with this article.
% \bibliographystyle{prd}
% \bibliography{ref} % if your bibtex file is called example.bib
\newpage
\twocolumngrid

\end{document}